\documentclass[12pt]{article}

\usepackage{a4wide}
\usepackage{epsfig}
\def\mb#1{\mbox{\scriptsize #1}}

\def\lsim{\mathrel{\rlap{\lower4pt\hbox{\hskip1pt$\sim$}}
    \raise1pt\hbox{$<$}}}                
\def\gsim{\mathrel{\rlap{\lower4pt\hbox{\hskip1pt$\sim$}}
    \raise1pt\hbox{$>$}}}                
\begin{document}
\vspace*{10mm}
\begin{center}  \begin{Large} \begin{bf}
A common scheme\\
for running NLO ep event generators\\
  \end{bf}  \end{Large}
  \vspace*{5mm}
  \begin{large}
Thomas Hadig$^a$, Gavin McCance$^b$\\
  \end{large}
\end{center}
$^a$ I.~Physikalisches.~Institut, RWTH Aachen, D-52056 Aachen, Germany\\
$^b$ ZEUS Collaboration,  mccance@zow.desy.de\\
\begin{quotation}
\noindent
{\bf Abstract:}
In this article we present a generic interface to several next-to-leading
order cross-section programs. This enables the user to implement his/her
code once and make cross-checks with different programs.
\end{quotation}
%

\section {Introduction}

Next-to-leading order QCD Monte Carlo programs are used for comparisons
of QCD perturbation theory and experimental data. Nowadays four programs
are available, which allow user defined observables; these are
MEPJET\cite{mepjet}, DISENT\cite{disent}, DISASTER++\cite{disaster}, and
JETVIP\footnote{JETVIP is currently not available in the common NLO
library.}\cite{jetvip}. With the availability of different programs
cross-checks become important. In order to make a comparison, the user
code has to be implemented for each program. This procedure is
susceptible to bugs and updating of several versions needs a decent
revision control system.

To eliminate this additional source of error, a common scheme was developed
and is presented here. The scheme consists of three independent parts,
which are described in the following sections. 

Section \ref{secsteer}
introduces the steering card. This file is read at the beginning of each
calculation and sets up the most important parameters. The user has the
ability to add additional parameters to steer his own user routines.

The interface to the user routines is described in section \ref{secuser}.
This code calculates the observables the user is interested in. For most
standard procedures, e.g.~performing boosts to different frames or
calculating the number of jets, special functions are available in a
library. These functions are explained in detail
in section \ref{seclib}.

A more detailed version of this manual including the full description
of the function library and an example are available with the source
code at {\tt www.desy.de/$\tilde{\ }$heramc/proceedings}
\cite{webversion}.

\section {Steering Card}
\label{secsteer}

At the beginning of the calculation a steering card is read from
standard input to set up the input parameters. This allows the easy
modification of the most important starting values even after
compilation.

While the generator's adjustable parameters are known, this is not true
for the user code. Therefore an easy interface is provided, which allows
the user to expand the set of steering card parameters.

\subsection{Basic Structure of the Steering Card}

The basic structure of the steering card is equivalent to that of most
Monte Carlo generators. The steering card consists of several
banks, labeled by a character string of four characters. The number of banks
in a steering card is not limited.

\setlength{\unitlength}{1cm}

Each bank consists of an unique four character label, an integer version
number and an arbitrary number of entries (including none).

An entry consists of an identifier and the value of the field. The value
can be of type integer, real or string, where string stands for a character
constant of arbitrary length.

Comments can be marked by an asterisk (*) in the first column or by an
exclamation mark (!) anywhere in a line. If a comment mark is found the
rest of the line is ignored.


\subsection{Predefined Banks}

Many parameters, like the particle density function and the number of
events, have to be set for all programs, these parameters are collected and
can be set by the use of a steering card bank called {\tt MOCA.} 

Below you see the complete {\tt MOCA} bank. Some of the entries are
described in detail below.

\begin{verbatim}
MOCA 0              ! NLO Monte Carlo steering card, Version 0
  'TYPE' 2          ! NLO program (1=Mepjet, 2=Disent, 3=Disaster++)
  'NEVE' 10000      ! Number of Events (required)
  'Q2MI' 10.        ! (D=100.) Q**2 min
  'Q2MA' 100.       ! (D=4000.) Q**2 max
  'XMIN' 0.         ! (D=0.) x min
  'XMAX' 1.         ! (D=1.) x max
  'XIMI' 0.         ! (D=0.) xi min
  'XIMA' 1.         ! (D=1.) xi max
  'YMIN' 0.         ! (D=0.) y min
  'YMAX' 0.7        ! (D=1.) y max
  'W2MI' 0.         ! (D=0.) Minimum W**2
  'W2MA' 100000.    ! (D=100000.) Maximum W**2
* PDFL and PDFH give the PDF libs used for LO (L,SET.eq.0) and 
* NLO (H,SET.ne.0) calculations, should be identical for MRS and 
* of corresponding type for CTEQ and GRV, 
* example : PDFL=5005 (GRV94 LO) and PDFH=5006 (GRV94 NLO) for
* MSbar scheme and PDFH=5007 (GRV94 NLO) for DIS scheme
  'PDFL' 3035       ! (D=3035=MRSH MSbar) PDF lib, 
  'PDFH' 3035       !   NGROUP*1000+NSET as in pdflib manual
  'ECMS' 90200.     ! (D=90200.=27.5*820.*4.) CMS energy
  'EP  ' 820.       ! (D=820) Proton energy
  'LEPT' 1          ! (D=1) incoming lepton (1: e-, 2: e+)
* The following values (upto * end) are given in the lab frame
  'ELMI' 11.        ! (D=0.) Minimum energy of scattered electron
* 'ELMA' 11.        ! (D=1000.) Maximum energy of scattered e-
  'TLMI' 150.       ! (D=0.) Minimum angle of scattered electron
  'TLMA' 172.5      ! (D=180.) Maximum angle of scattered e-
* end
  'NFL ' 5          ! (D=5) number of flavours
  'SFQ2' 1.         ! (D=1.) factorization scale factor for Q**2
  'SRQ2' 1.         ! (D=1.) renormalization scale factor for Q**2
  'SFKT' 0.         ! (D=0.) factorization scale factor for kt**2
  'SRKT' 0.         ! (D=0.) renormalization scale factor for kt**2
  'SFPT' 0.         ! (D=0.) factorization scale factor for pt**2
  'SRPT' 0.         ! (D=0.) renormalization scale factor for pt**2
  'SFCO' 0.         ! (D=0.) factorization scale factor for 1
  'SRCO' 0.         ! (D=0.) renormalization scale factor for 1
  'IAEL' 0          ! (D=0) switch for alpha_electromagnetic
                    ! 0 : fixed, 1 : running
  'AELM' 0.00729735308 ! (D=0.00729735308) alpha_electromagnetic
  'MASS' 0.15       ! (D=0.15) mass of strange quark in GeV
  'MASC' 1.4        ! (D=1.4) mass of charm quark in GeV
  'MASB' 4.4        ! (D=4.4) mass of bottom quark in GeV
  'MAST' 174        ! (D=174) mass of top quark in GeV
  'ALOO' 2          ! (D=2) number of loops for alpha_s running,
                    ! -1: pdf routine, 0: fixed, 1|2|3: 1|2|3 loop 
  'LAM4' 0.3        ! (D=0.3) Lambda_4_MSbar for running alphas
                    ! or alphas for fixed alphas
\end{verbatim}

\begin{itemize}
\item {\tt ITYPE} is coded as follows~:
\begin{description}
\item[ITYPE=0~:] unknown
\item[ITYPE=1~:] MEPJET
\item[ITYPE=2~:] DISENT
\item[ITYPE=3~:] DISASTER++
\end{description}

\item {\tt NEVE} is the only field required for each run. It gives the 
number of events to be generated\footnote{It should be noted for
DISASTER++, however, that the product FFIN * POINTS determines the
number of events generated. Using the library default values,
\mbox{DISASTER++} will produce 50\% more events than DISENT.}.

\item {\tt PDFL} and {\tt PDFH} specify the parton density functions 
used. Two values are given in order to enable the distinction of leading
order and next-to-leading order processes. The format of each of the
values is {\tt NGROUP * 1000 + NSET} where {\tt NGROUP} and {\tt NSET}
are given by the PDFLIB.
\item The renormalization and factorization scale can be set using the
formula $$\mu_s^2 = f_{Q2} Q^2 + f_{PT} p_t^2 + f_{KT} k_t^2 + f_{CO}$$
where the factors $f_{xy}$ are set with the steering parameter {\tt Ssxy.}
\item {\tt IAEL} and {\tt AELM} steer the running of $\alpha_{\mb{el.mag.}}$
and {\tt ALOO} and {\tt LAM4} steer the running of $\alpha_s$
accordingly. When {\tt ALOO} is set to -1, the $\alpha_s$ calculation included
in the pdf library is used. If it is 0, $\alpha_s\equiv${\tt LAM4} for every
scale, otherwise {\tt LAM4} corresponds to $\Lambda^4_{\overline{MS}}$ and
the masses {\tt MASx} are used for calculating 
$\Lambda^{(3,5,6)}_{\overline{MS}}.$
\end{itemize}

Some parameters are still program specific, therefore for each generator an
additional steering card exists. Below you can find the steering card of
DISASTER++\cite{disaster} and DISENT\cite{disent}.

\begin{verbatim}
DISE 0              ! DISENT specific steering card
* 'SEDL' -1         ! (12345) Lower seed value for random generator
                    ! if set to negative value a time-depended
                    !  value will be used
* 'SEDH' -1         ! (678900) Higher seed value for random generator
                    ! if set to negative value a process-depended
                    !  value will be used
  'SCHE' 0          ! (D=0) 0 : MSbar, 1 : DIS
                    ! Please note : the PDF has to be chosen equivalent
  'NPO1' 2          ! (D=2) The parameters NPO1 and NPO2 are internal
  'NPO2' 4          ! (D=4) parameters used by Disent, please refer
                    ! to the program manual
DISA 0              ! DISASTER++ specific steering card
  'BORN' 2          ! (D=2) number of particles in born term
  'PROC' 0          ! (D=1) order of process (0 : LO, 1 : NLO)
  'FPRE' 1.5        ! (D=1.5) factor for # points in preparation run
  'FFIN' 1.5        ! (D=1.5) factor for # points in final run

MEPJ 0              ! MEPJET specific steering card
  'BOSO' 1          ! exchange boson (1:gamma, 2:Z, 3:gammaZ, 4:W)
  'BORN' 2          ! (D=2) number of particles in born term
  'PROC' 0          ! (D=1) order of process (0 : LO, 1 : NLO)
  'NPRO' 100        ! (D=100) process number (quark+gluon, unpolarized)
                    ! see Mepjet manual for details (iproc)
  'IMAS'   0        ! (D=0) massive calculation (0=no, 1=yes)
  'IPDF'   4        ! (D=4) pdf crossing function
  'ITER'   4        ! (D=4) number of iterations for vegas adaptation
  'SMIN'   0.1      ! (D=0.1) smin cone size
  'PTIN'   0.1      ! (D=0.1) minimal p_t in lab frame for partons
  'AMIN' -10.       ! (D=-10) minimal pseudo rapidity in lab for partons
  'AMAX'  10.       ! (D= 10) maximal pseudo rapidity in lab for partons
  'FILE' 'mepjet'   ! base name for vegas grid files
\end{verbatim}

Please note, that due to the usage of crossing functions the parton density
functions for Mepjet is not set with {\tt PDFL, PDFH,} but by the {\tt IPDF}
switch using the crossing functions found in the {\tt pdfcross} directory.


\subsection{User Functions for Steering Card Access}

In order to add an additional parameter to the steering card, two steps are
required. First the parameter has to be initialized before the steering
card is read using one of the following routines~:

\begin{itemize}
\item {\tt call SCARINT(bank,identifier,value)} for integer parameters
\item {\tt call SCARREAL(bank,identifier,value)} for double precision  parameters
\item {\tt call SCARCHAR(bank,identifier,value)} for string parameters
\end{itemize}
where {\tt bank} is the four letter bank name, {\tt identifier} is the one
to four character identifier and the {\tt value} is the default value of the
type corresponding to the type of the parameter.

After the steering card has been read, the user can retrieve the current
parameter value. To do so, three functions are available~:

\begin{itemize}
\item {\tt value = GCARINT(bank,identifier)} for integer parameters
\item {\tt value = GCARREAL(bank,identifier)} for double precision parameters
\item {\tt value = GCARCHAR(bank,identifier)} for string parameters
\end{itemize}
where {\tt bank} and {\tt identifier} correspond to the values given above
and {\tt value} will be the one given in the steering card or the default
parameter, if the corresponding bank or identifier in the bank was not set
in the steering card. Values and banks may appear more than once in a
steering card, but only the last value is stored and can be retrieved.

In principle the get routines could be called each time one of the values
is needed, but since this requires several string comparisons, it is
rather slow. The recommended procedure is to get the values once and store
them in a FORTRAN common block.



\section {User Routines}
\label{secuser}

In this section, the routines that have to be supplied by the user in order
to run the programs are described. Therefore the general scheme is
sketched here. The number of user routines is quite large, but most
routines can be replaced by empty or short routines if no special action
is needed.

\begin{figure}[htb]
\begin{picture}(11,12)
\put(1,11.5){\line(0,-1){11.7}}
\put(1,11.05){\line(1,0){0.3}}
\put(1.5,11){ {\tt disinit} }
\put(1,10.05){\line(1,0){0.3}}
\put(1.5,10){ {readsteer} }
\put(1,9.05){\line(1,0){0.3}}
\put(1.5,9){ {\tt init} }

\put(1,8.05){\line(1,0){0.6}}
\put(1.6,8.05){\line(0,-1){1.3}}
\put(1.7,8){ {\it adaptation loop} }
\put(1.6,7.05){\line(1,0){0.3}}
\put(2.1,7){ {\tt disphase} }

\put(1,6.05){\line(1,0){0.6}}
\put(1.6,6.05){\line(0,-1){5.3}}
\put(1.7,6){ {\it event loop} }
\put(1.6,5.05){\line(1,0){0.3}}
\put(2.1,5){ {\tt evtinit} }
\put(1.6,4.05){\line(1,0){0.6}}
\put(2.2,4.05){\line(0,-1){2.3}}
\put(2.3,4){ {\it contribution loop} }
\put(2.2,3.05){\line(1,0){0.3}}
\put(2.6,3){ {\tt disphase} }
\put(2.2,2.05){\line(1,0){0.3}}
\put(2.6,2){ {\tt discontr} }
\put(1.6,1.05){\line(1,0){0.3}}
\put(2.1,1){ {\tt evtterm} }
\put(1,0.05){\line(1,0){0.3}}
\put(1.5,0){ {\tt disterm} }
\end{picture}
\caption{\label{fig_overview}{\it Overview of the calling tree.}}
\end{figure}
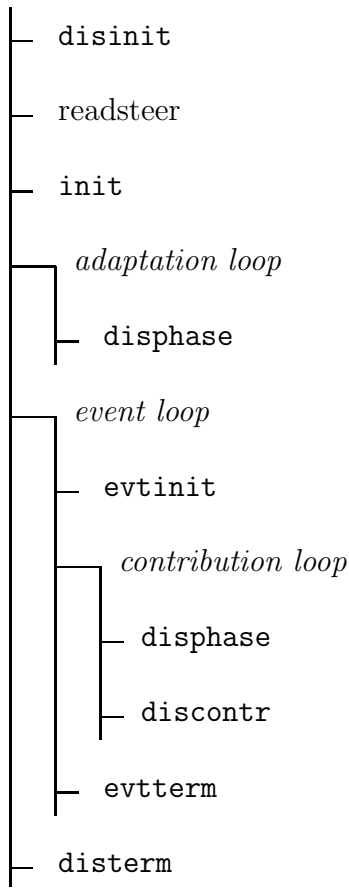

Figure \ref{fig_overview} shows the calling tree of the relevant functions
and subroutines. Routines printed in {\tt typewriter} have to be supplied
by the user and are described in detail below, functions in {\it italic}
and roman are provided by the corresponding cross-section Monte Carlo and
the library, respectively.

\subsection{Initialization and Termination Routines}

Several initialization and termination routines are called to allow the
user to predefine values, to open files, or to book histograms. None of
these subroutines take an argument. In the following each routine is
described.

{\tt disinit} is called only once at the beginning of the program. This is
the place to initialize the users steering card values.

After the steering card has been read, {\tt init} is called. Here the
parameter values can be copied to a common block, as proposed at the end
of section \ref{secsteer}. Also output files should be opened here.

Corresponding to these initialization routines a termination routine {\tt
disterm} exists. This function is called once at the end of the run. Here
all files should be closed.

In addition to these called-once routines an additional set of init and
term routines exists. The {\tt evtinit} subroutine is called before a new
event is generated and {\tt evtterm} afterwards. Since all contributions to
the observable of one event have to be added before the error can be
calculated\cite{disent_manual}, the main summation can only be done in
those routines. Please note that the {\tt evtterm} routine will not
be called if an event has to be dropped due to technical reasons (e.g. after
a floating point exception). See the web version of this 
manual \cite{webversion} for an example.
 

\subsection{Main User Routines}

For each event a number of phase spaces is generated, each of these phase
spaces can have several contributions. The main idea is to calculate your
observables once for each phase space {\tt iphasno} and add all weights of
the corresponding contributions to the event weight. The calculation of the
observables is done in {\tt disphase} and the summing is done in {\tt
discontr.} The user has to take care that the observables are saved in a
common block and are available when {\tt discontr} is called. It is
sufficient to allocate space for up to 30 different phase spaces.

In addition to the actual event sampling, some programs perform an
adaptation loop to adapt the phase space to the region that is actually used. In
this additional loop {\tt disphase} is called only. The adaptation is then
done according to the return value.

The syntax is
\begin{verbatim}
      function disphase (iphasno, npar, ps, iadapt)

      double precision disphase
      integer          iphasno, npar, iadapt
      double precision ps(4,10)
\end{verbatim}
and      
\begin{verbatim}
      subroutine discontr (iphasno, ntype, nalp, nalps, n2pi,
     +   ilognf, jmin, jmax, weight )
      
      integer          iphasno, ntype, nalp, nalps, n2pi
      integer          ilognf, jmin, jmax 
      double precision weight(-13:11)
\end{verbatim}
where the arguments have the following meaning~:
\begin{itemize}
\item {\tt iphasno} gives the number of the phase space.
\item {\tt npar} is the number of outgoing particles (excluding the scattered
lepton)
\item {\tt ntype} is the type of the contribution
 \begin{description}
  \item[ntype = -1] unknown
  \item[ntype = 0] tree level
  \item[ntype = 1] subtraction counterterm
  \item[ntype = 2] finite virtual term
  \item[ntype = 3] finite collinear term
  \item[ntype = 4] renormalization term
 \end{description}
\item {\tt nalp} and {\tt nalps} are the orders of $\alpha_{\mb{elm}}$
and $\alpha_{s},$ respectively. {\tt n2pi} gives the power of an additional
factor of $\displaystyle{1\over 2\pi}.$
\item {\tt ilognf} selects one of the following factors~:
 \begin{description}
 \item[ilognf=0~:] $\lambda=1$
 \item[ilognf=1~:] $\displaystyle{\lambda=\log\left( {\mu_r^2\over Q^2}\right)}$
 \item[ilognf=2~:] $\displaystyle{\lambda=N_f\log\left( {\mu_r^2\over Q^2}\right)}$
 \item[ilognf=3~:] $\displaystyle{\lambda=\log\left( {\mu_f^2\over Q^2}\right)}$
 \item[ilognf=4~:] $\displaystyle{\lambda=N_f\log\left( {\mu_f^2\over Q^2}\right)}$
 \end{description}
\item {\tt ps($i,j$)} defines the phase space. The array contains several
particles distinguished by the second index $j.$
 \begin{description}
  \item[$j=1~:$] incoming proton
  \item[$j=2~:$] incoming lepton
  \item[$j=3~:$] boson
  \item[$j=4~:$] outgoing lepton
  \item[$j=5~:$] outgoing proton remnant
  \item[$j=6~:$] incoming parton
  \item[$j=7,\dots,10~:$] outgoing partons
 \end{description}
The first index gives the 4-vector components, where $i =1$ corresponds to the
energy, and $i=2,3,4$ to the $x,y,$ and $z$ components of the momentum,
respectively.
\item {\tt iadapt} is set to $1$ during the adaptation loop. The value returned
by {\tt disphase} is used to adapt the integration and sampling phase space.
If a negative is returned, a default adaptation calculation is used.
\item {\tt weight} is an array containing several weights. For each 
contribution only a specific range in this array, defined by {\tt jmin}
and {\tt jmax} is valid. Each weight has to be multiplied by a factor $\rho_i$
and then all weights have to be summed. The factor $\rho_i$
for the weight $i$ depends on the particle density functions $f_{\alpha}$
for the different flavours $\alpha$ and is calculated by~: 
 \begin{description}
 \item[$i=-8\dots -13~:$] $f_{\bar \alpha}$ with $\alpha = u(-8), d, s, c, b, t(-13).$
 \item[$i=-7~:$] $f_g$
 \item[$i=-1\dots -6~:$] $f_{\alpha}$ with $\alpha = u(-6), d, s, c, b, t(-1).$
 \item[$i=0~:$] $1$
 \item[$i=1,4~:$] $\displaystyle{\sum_{\alpha=1}^{N_f} Q_{\alpha}^2 f_{\alpha}}$
 \item[$i=2,5~:$] $\displaystyle{\sum_{\alpha=1}^{N_f} Q_{\alpha}^2 f_{\bar \alpha}}$
 \item[$i=3~:$] $\displaystyle{\sum_{\alpha=1}^{N_f} Q_{\alpha}^2 f_g}$
 \item[$i=6~:$] $\displaystyle{(1- N_f)\sum_{\alpha=1}^{N_f} Q_{\alpha}^2 f_{\alpha}}$
 \item[$i=7~:$] $\displaystyle{(1- N_f)\sum_{\alpha=1}^{N_f} Q_{\alpha}^2 f_{\bar \alpha}}$
 \item[$i=8~:$] $\displaystyle{\sum_{\alpha=1}^{N_f}f_{\alpha}
               \sum_{\beta=1,\beta\ne\alpha}^{N_f} Q_{\beta}^2}$
 \item[$i=9~:$] $\displaystyle{\sum_{\alpha=1}^{N_f}f_{\bar\alpha}
               \sum_{\beta=1,\beta\ne\alpha}^{N_f} Q_{\beta}^2}$
 \item[$i=10~:$] $\displaystyle{\sum_{\alpha=1}^{N_f}f_{\alpha}Q_{\alpha}
               \sum_{\beta=1,\beta\ne\alpha}^{N_f} Q_{\beta}}$
 \item[$i=11~:$] $\displaystyle{\sum_{\alpha=1}^{N_f}f_{\alpha}Q_{\bar\alpha}
               \sum_{\beta=1,\beta\ne\alpha}^{N_f} Q_{\beta}}$
 \end{description}
\end{itemize}

The full weight of one contribution is then
$$w_{\mb{contribution}} =
 \lambda \alpha_s^{\mb{\tt ialps}} \alpha_{\mb{elm}}^{\mb{\tt ialp}}
 \left({1\over 2\pi}\right)^{\mb{\tt n2pi}}
 \sum_{i= \mb{\tt jmin}}^{\mb{\tt jmax}} {\tt weight}(i) \rho_i      $$

\noindent
This somewhat complicated procedure can be performed by a library
function {\tt double wsum (nord, nalp, nalps, n2pi, ilognf, jmin, jmax,
weight, fscale, rscale, alphas).}\\
See section 2 of the web 
version\cite{webversion} for an example.
{\tt nord} selects the
leading or higher order particle density function (see steering card values
{\tt PDFL} and {\tt PDFH}). {\tt fscale} and {\tt rscale} are the input
values for the factorization and renormalization scales, respectively.
{\tt alphas} is an output parameter returning the value of $\alpha_s$ at
this phase space point.

{\small
Additional information for experienced users~:
For DISASTER++ all phase spaces for one event are generated at the
beginning of the event and the {\tt disphase} routine is called for each
phase space, before the first call to {\tt discontr} is made. The {\tt ntype}
information is 0 for tree level and -1 otherwise.
In DISENT the sequence is different, for each contribution {\tt disphase} 
and {\tt discontr} are called right after each other. The {\tt ntype} 
information is fully available. 
The adaptation loop is performed in DISASTER++, only.
}


\subsection{Common Blocks}

Some common blocks are provided to the {\tt disphase} and {\tt discontr}
routines. They contain the steering card values for the predefined bank and
the event kinematics.

Here are the definitions\footnote{For a detailed discussion see section
\ref{secsteer}}~:

\begin{verbatim}
      INTEGER           MOCAVERS, NEV, ITYPE, IALELM, NPDFL, NPDFH
      INTEGER           NFL, ALOOP, LEPTON
      DOUBLE PRECISION  Q2MIN, Q2MAX, XMIN, XMAX, YMIN, YMAX, S
      DOUBLE PRECISION  ELMIN, ELMAX, TLMIN, TLMAX, W2MIN, W2MAX
      DOUBLE PRECISION  SRQ2, SFQ2, SRPT, SFPT, SRKT, SFKT, SRCO
      DOUBLE PRECISION  SFCO, DEFAEM, XIMIN, XIMAX, EP
      DOUBLE PRECISION  LAMBDA3, LAMBDA4, LAMBDA5, LAMBDA6
      DOUBLE PRECISION  MASSS, MASSC, MASSB, MASST, LEPTON

      COMMON /STERMOCA/MOCAVERS, NEV, ITYPE, IALELM, DEFAEM, 
     +  Q2MIN, Q2MAX, XMIN, XMAX,YMIN,YMAX, S, EP, NPDFL, 
     +  NPDFH, ELMIN, ELMAX, TLMIN, TLMAX, W2MIN, W2MAX, 
     +  SRQ2, SFQ2, SRPT, SFPT, SRKT, SFKT, SRCO, SFCO,
     +  XIMIN, XIMAX, NFL, ALOOP,
     +  LAMBDA3, LAMBDA4, LAMBDA5, LAMBDA6,
     +  MASSS, MASSC, MASSB, MASST
\end{verbatim}
and 
\begin{verbatim}
      INTEGER          DISEVERS, ISEEDL, ISEEDH, NSCHEME
      INTEGER          NPO1, NPO2

      COMMON /STERDISE/DISEVERS, ISEEDL, ISEEDH, NSCHEME, 
     +  NPO1, NPO2


      INTEGER          DISAVERS, IPROC, IBORN
      DOUBLE PRECISION FPRE, FFIN

      COMMON /STERDISA/DISAVERS, IPROC, FPRE, FFIN, IBORN


      INTEGER MEPJVERS, IMBOSO, IMPROC, IMMASS, IMPDF,
     +        IMBORN, IMEPROC, IMITER
      DOUBLE PRECISION RMSMIN, PTMIN_DEF, YMIN_DEF, YMAX_DEF
      COMMON /STERMEPJ/MEPJVERS, IMBOSO,
     +                IMBORN, IMEPROC, IMITER, IMPDF,
     +                RMSMIN, IMPROC,  IMMASS,
     +                PTMIN_DEF, YMIN_DEF, YMAX_DEF
\end{verbatim}
where the values correspond to the steering card parameters.

The additional {\tt KINE} common block defines the event kinematics with
some invariants and some energies in the laboratory frame of reference.

\begin{verbatim}
      DOUBLE PRECISION Q2, XB, YB, W2, SUMKT2, SUMPT2,
     +                 ESCELE, THSCELE, EE, XI, ALPHA

      COMMON /KINE/EE, XI, Q2, XB, YB, W2, ESCELE, THSCELE, 
     +   ALPHA, SUMKT2, SUMPT2
\end{verbatim}



\section {Interface to HzTool}
\label{sechztool}

HzTool\cite{hztool} is a package that provides code to compare data
plots published by the HERA experiments to Monte Carlo programs. Some of
the plots could also be compared to next-to-leading order calculations
and an interface from this library to HzTool is presented here.

Note that several observables are not applicable in next-to-leading
order programs, such as particle or track multiplicities, and others
would require to add hadronisation effects to the calculation. Therefore
the comparison is a priori limited to a few observables.

HzTool reads data from the HEPEVT common block, a standard format for
high energy physics Monte Carlo programs. In addition, HzTool expects
that every call contains the full information for one event with a
corresponding weight. Since different contributions to one event have to
use a special error treatment, a decent error calculation without
changing all HzTool routines is not possible.

The restrictions implied are therefore:
\begin{itemize}
\item Only observables available to next-to-leading programs allow 
   comparisons.
\item Be especially aware of cuts that spoil cancelations of 
 divergencies\cite{durham}.
\item Errors calculated by HzTool are not valid.
\item Differences according to non-perturbative effects can be 
  expected.
\item Some NLO programs might give unusable results, e.g.\ dijet rates 
  need the simultaneous calculation of ${\cal{O}}(1)$ and
  ${\cal{O}}(\alpha_s)$ tree level diagrams, which is e.g.\ not possible
  in Disaster++.
\end{itemize}

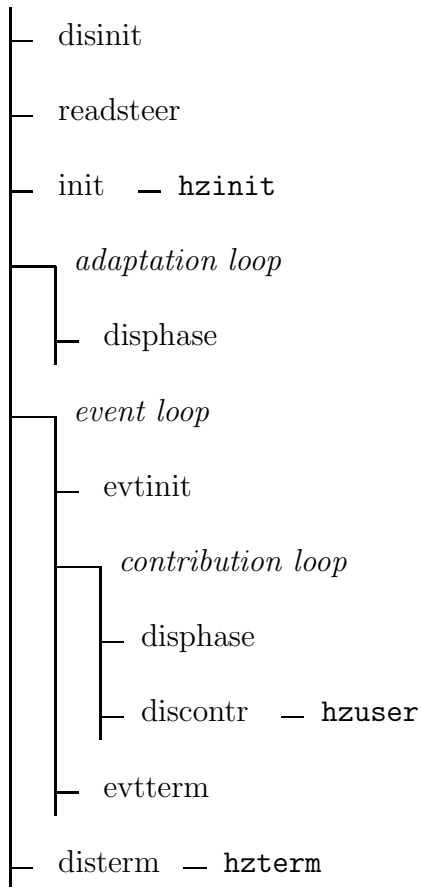
\begin{figure}[htb]
\begin{picture}(11,12)
\put(1,11.5){\line(0,-1){11.7}}
\put(1,11.05){\line(1,0){0.3}}
\put(1.5,11){ {disinit} }
\put(1,10.05){\line(1,0){0.3}}
\put(1.5,10){ {readsteer} }
\put(1,9.05){\line(1,0){0.3}}
\put(1.5,9){ {init} }

\put(2.7,9.05){\line(1,0){0.3}}
\put(3.1,9){ {\tt hzinit} }

\put(1,8.05){\line(1,0){0.6}}
\put(1.6,8.05){\line(0,-1){1.3}}
\put(1.7,8){ {\it adaptation loop} }
\put(1.6,7.05){\line(1,0){0.3}}
\put(2.1,7){ {disphase} }

\put(1,6.05){\line(1,0){0.6}}
\put(1.6,6.05){\line(0,-1){5.3}}
\put(1.7,6){ {\it event loop} }
\put(1.6,5.05){\line(1,0){0.3}}
\put(2.1,5){ {evtinit} }
\put(1.6,4.05){\line(1,0){0.6}}
\put(2.2,4.05){\line(0,-1){2.3}}
\put(2.3,4){ {\it contribution loop} }
\put(2.2,3.05){\line(1,0){0.3}}
\put(2.6,3){ {disphase} }
\put(2.2,2.05){\line(1,0){0.3}}
\put(2.6,2){ {discontr} }

\put(4.6,2.05){\line(1,0){0.3}}
\put(5.0,2){ {\tt hzuser} }

\put(1.6,1.05){\line(1,0){0.3}}
\put(2.1,1){ {evtterm} }
\put(1,0.05){\line(1,0){0.3}}
\put(1.5,0){ {disterm} }

\put(3.3,0.05){\line(1,0){0.3}}
\put(3.7,0){ {\tt hzterm} }
\end{picture}
\caption{\label{fig_hztool}{\it Overview of the calling tree when using HzTool.}}
\end{figure}

To use the HzTool interface, add the file nlolib/src/hztool/hzhep.f to 
your source. This routine provides implementations to the standard user
routines and therefore replaces the interface specified in section
\ref{secuser} and figure \ref{fig_overview} by the one given in figure
\ref{fig_hztool}.

The routines that have to be provided by the user are {\tt hzinit(),
hzterm(),} and {\tt hzuser()}. These routines should call the {\tt
hz{\it xxxxx}} routines with {\tt iflag} equal to 1, 3, and 2
respectively. See the HzTool manual \cite{hztool} for more information.

\subsection{Note for authors of {\tt hz{\it xxxxx}} routines}

In order to work with this library, {\tt hz{\it xxxxx}} routines have to
fulfil some additional requirements.

\begin{itemize}
\item Add the common block {\tt HZNLO} to your routine.
\item Initialize the variable {\tt NLO} to 0 in the {\tt iflag.eq.1}
  section.
\item Add an {\tt if} clause to every hbook fill or weight summation 
  line in order to check, whether the current event counts for the 
  desired process, e.g.\ check for total cross sections, whether the 
  process is a ${\cal{O}}(1)$ born term or a ${\cal{O}}(\alpha_s)$ 
  correction. This check can easily be done using the variables defined 
  in the {\tt HZNLO} common block.
\end{itemize}

The {\tt HZNLO} common block is defined as follows:
\begin{verbatim}
      INTEGER NLO, TOT, DIJET, TRIPJET, QUADJET
      COMMON /HZNLO/NLO, TOT, DIJET, TRIPJET, QUADJET
\end{verbatim}

An example might look like:
\begin{verbatim}
      IF (iflag.eq.1) THEN
C... init step
        NLO=0
      ELSE IF (iflag.eq.2) THEN
C... fill step

C... make some phase space cuts

CTH add total x-section
        IF ((NLO.eq.1).AND.(TOT.ne.1)) GOTO 11112
         nall=nall+xw
11112   CONTINUE

C... run some jet algorithm and make some cuts on dijet events

        IF (twojet) THEN
          IF ((NLO.eq.1).AND.(DIJET.ne.1)) GOTO 11114
            nall2=nall2+xw
            call hfill(121,x,0.,xw)
            call hfill(131,q2,0.,xw)
11114     CONTINUE
        ENDIF
      ENDIF
\end{verbatim}


\section {Function Library}
\label{seclib}

In addition to the unified interface a set of subroutines is defined.
Please, see the full list below.

\begin{itemize}
 \item jet algorithms
  \begin{itemize}
   \item JADE
   \item $k_t$
   \item longitudinal boost-invariant $k_t$
  \end{itemize}
 \item event shape variables
 \item Lorentz boosts
 \item support routines for calculation of momentum, angles, masses, etc.
\end{itemize}

\noindent
For the full documentation of most routines see the web 
version\cite{webversion}.

\subsection{Jet Algorithms}

Several jet algorithms exist. The calling sequence for those is
\begin{verbatim}
      double precision P(4,*), YCUT, SCALE, VEC(4,*)
      integer          NPA, IRECOM, NUM

      call <algo>jet(P,NPA,YCUT,SCALE,I,IRECOM,NUM,VEC)
\end{verbatim}
where {\tt <algo>} is replaced by the name of the algorithm (i.e. one of
{\tt jade, kt}).

The common arguments are~:
\begin{itemize}
\item {\bf P} the input particles, 
\item {\bf NPA} the number of input particles, 
\item {\bf VEC} the output jet axes and energies,
\item {\bf NUM} the number of jets and
\item {\bf IRECOM} gives the recombination scheme for two particles (see
the description of the support routine {\tt vadd} in section \ref{sec_support}
for more details).
\end{itemize}

The remaining arguments are algorithm dependent. Please refer to the
individual manual. The basic idea is, that {\bf SCALE} corresponds to a
scaling factor and {\bf YCUT} corresponds to a cut value. This is for
example the scale for the invariant mass and the maximum mass over scale
value for the jade algorithm and the minimum jet $E_t$ and the maximal cone
radius for the cone algorithm. {\bf I} is an additional input value with no
predefined meaning.

For the {\tt KTCLUS} package, a slightly different calling sequence is used~:
\begin{verbatim}
      SUBROUTINE KTINCJET(P,NPA,PTMIN,NUM,V)

      DOUBLE PRECISION P(4,*),V(4,*),PTMIN
      INTEGER NUM,NPA

      SUBROUTINE KTCLUSJET(P,NPA,SCALE,YCUT,NUM,V)

      DOUBLE PRECISION P(4,*),V(4,*),SCALE,YCUT
      INTEGER NUM,NPA
\end{verbatim}

where the arguments correspond to the arguments explained in the beginning
of the section.


\subsection{Lorentz Boost}

Two routines are provided to perform Lorentz boosts to other reference
frames.

\begin{verbatim}
      SUBROUTINE boost1 (V,BR,BZ,BXZ,IDIR,IERR)

      DOUBLE PRECISION V(4),BR(4),BZ(4),BXZ(4)
      INTEGER          IDIR,IERR

      SUBROUTINE boost (P,V,N,BR,BZ,BXZ,IDIR,IERR)

      DOUBLE PRECISION P(4,*),V(4,*),BR(4),BZ(4),BXZ(4)
      INTEGER          N,IDIR,IERR
\end{verbatim}

{\tt boost1} boosts one vector, given in array {\tt V} into the new frame,
where {\tt V} will be used for input and output.
{\tt boost} can boost {\tt N} vectors given in array {\tt P} into the new
frame. Here the original vectors are conserved and the new vectors are
filled in array {\tt V.} {\tt boost} should be preferred, when more than
one vector is to be boosted, since the calculation of the rotation matrices
is done only once for all vectors.

The boost is specified by three 4-vectors, which will have the following
characteristics in the boosted frame~: {\tt BR} will be the 0-vector,
{\tt BZ} will point in z-direction and {\tt BXZ} will lie in the x-z-plane.
If {\tt IDIR} is zero a normal boost will be performed, if {\tt IDIR} is
one, the boost is reverted such that boosting the returned particles in 
{\tt V} with the given boost vectors (and {\tt IDIR=0}) would result in
the original particles {\tt P.}

If an error occurs during boosting {\tt IERR} will contain a non-zero value.

The routines
\begin{verbatim}
      SUBROUTINE blab2br1 (V,IERR)
      SUBROUTINE blab2br (P,O,N,IERR)
      SUBROUTINE bbr2lab1 (V,IERR)
      SUBROUTINE bbr2lab (P,O,N,IERR)

      DOUBLE PRECISION P(4,*),O(4,*),V(4)
      INTEGER          N,IERR
\end{verbatim}

perform the boosting of one or several vectors from the H1 laboratory frame to
the Breit frame and vice versa.
 

\subsection{Event Shape Routines}

A single routine is provided to calculate a variety of event shape variables.

\begin{verbatim}
      SUBROUTINE getevtshape (NPAR,PS,SCALE,SHAPE,SUME)

      DOUBLE PRECISION PS(4,10),SCALE,SHAPE(NOVAR),SUME
      INTEGER          NPAR, NOVAR
      PARAMETER       (NOVAR=20)   ! the current number of variables

\end{verbatim}

where the input parameters are

\begin{itemize}
\item {\bf NPAR} the number of input particles, 
\item {\bf PS(4,11)} the four vectors {\tt PS()} from {\tt disphase}, 
\item {\bf SCALE} the scale to normalise to for some variables, usually $Q$.
\end{itemize}

the subroutine outputs

\begin{itemize}
\item {\bf SUME} the total energy in the current region of the Breit frame, 
\item {\bf SHAPE(NOVAR)} and array with all the event shapes. They are indexed by a letter code eg. {\tt SHAPE(itz)} for current jet thrust. 
\end{itemize}

The table below gives the indexes for all 14 of the available event shapes. The remaining 6 variables in {\tt SHAPE(20)} are internal or obsolete and should not be used.

\begin{table}[htb]
\begin{center}
\begin{tabular}{|l|c|}
\hline
{\tt SHAPE(itz)}  & Current jet thrust\\ \hline
{\tt SHAPE(itz2)} & 2nd moment of current jet thrust\\ \hline
{\tt SHAPE(itm)}  & Thrust axis thrust\\ \hline
{\tt SHAPE(itm2)} & 2nd moment of thrust axis thrust\\ \hline
{\tt SHAPE(ijb)}  & Current jet broadening\\ \hline
{\tt SHAPE(ijb2)} & 2nd moment of current jet broadening\\ \hline
{\tt SHAPE(ijm)}  & Current jet mass\\ \hline
{\tt SHAPE(ic4)}  & C-parameter (normalised to energy)\\ \hline
{\tt SHAPE(ic5)}  & C-parameter (normalised to momentum)\\ \hline
{\tt SHAPE(ic6)}  & C-parameter (normalised to {\tt $scale/2$})\\ \hline
{\tt SHAPE(ied)}  & Energy deficit\\ \hline
{\tt SHAPE(iob)}  & Oblateness\\ \hline
{\tt SHAPE(itr)}  & Thrust to resultant axis\\ \hline
{\tt SHAPE(ijbr)} & Jet broadening to resultant axis\\ \hline
\hline
\end{tabular}
\end{center}
\end{table}

The index variables are available by {\tt including} the file {\tt index.inc}. The routine is designed to be called from within {\tt disphase} and passed the common library four-vectors {\tt ps()} directly. No cut on the total energy in the current region (to ensure infrared safety) is performed within the routine; this quantity is returned and any cut is the responsibility of the user routine.

\subsection {Support Routines}
\label{sec_support}

A collection of small functions are available. Most of the routines come
in two versions, one that takes a vector and one that takes an array of
vectors and an index as arguments.

\begin{itemize}
\item {\tt vcopy (A,I,B,J)} copies the i'th vector of array A to the j'th
vector of array B.
\item {\tt vadd(A,I,B,J,IRECOM)} adds the j'th vector in array B to the
i'th vector in array A. The available recombination schemes are~:
 \begin{description}
 \item[{\tt IRECOM=1}] E/JADE : ${\bf p} := {\bf p_a} + {\bf p_b.}$
 \item[{\tt IRECOM=2}] E0 : $E := E_a+E_b, \vec p := {|\vec p|\over E}
    (\vec p_a + \vec p_b)$ (momentum rescaling)
 \item[{\tt IRECOM=3}] P : $\vec p :=  (\vec p_a + \vec p_b), E = |\vec p|$
    (energy rescaling)
 \item[{\tt IRECOM=4}] Snowmass : $p_t = p_{t,a} + p_{t,b}, 
  \eta = {1\over p_t} ( \eta_a p_{t,a} + \eta_b p_{t,b}),
  \varphi = {1\over p_t} ( \varphi_a p_{t,a} + \varphi_b p_{t,b}),
  E = |\vec p|$
 \end{description}
\item{\tt P = ATH (A,I)} calculates the theta angle (in radian) of the  i'th
  vector of array A wrt. the $+z$-axis.
\item{\tt P = VP2 (V)} and {\tt P = AP2(A,I)} returns the momentum squared
    of the particle V and A(i), respectively.
\item{\tt P = VMASS2 (V)} and {\tt P = AMASS2(A,I)} returns the invariant
    mass squared of the particle V and A(i), respectively.
\end{itemize}

In the above functions and subroutines the variables are defined as follows~:
\begin{verbatim}
      DOUBLE PRECISION A(4,*), B(4,*), V(4), P
      INTEGER          I,J,IRECOM
\end{verbatim}


\section {Summary}

In this paper we have presented a scheme for a complete library of
next-to-leading order QCD programs. This library consists of three parts; a
steering card mechanism, a unified interface for the user routines and an
expandable set of FORTRAN library functions.


\section*{Acknowledgments}

We would like to thank the authors for their support and appreciate
the comments and suggestions on the program and this paper. We also
thank the organizers and conveners of this workshop for the interesting
topics.



\end{document}